\documentstyle[12pt]{article}
\begin{document}
{\large {\bf Global Anomalies in the Batalin Vilkovisky
Quantization}}\\
\vspace{1cm}

{\large Ricardo Amorim\footnote{\noindent e-mail: amorim @ if.ufrj.br} and
Nelson R. F. Braga\footnote{\noindent e-mail: braga @ if.ufrj.br}}\\
\vspace{1cm}

Instituto de F\'\i sica, Universidade Federal  do Rio de Janeiro,\\
Caixa Postal 68528, 21945  Rio de Janeiro,
RJ, Brazil\\
 
\vspace{0.5cm}
\abstract
The  Batalin  Vilkovisky (BV) quantization provides a general procedure for calculating anomalies associated to gauge   symmetries.  Recent results show that even higher loop order contributions can be calculated by introducing an appropriate regularization-renormalization scheme.
However, in its standard form, the BV quantization is not
sensible to quantum violations of the classical conservation of  Noether currents, the so called global anomalies. 
We show here that the BV field antifield method can be extended in such a way that  the Ward identities involving  divergencies of global Abelian currents can be calculated from the generating functional, a result that would not be obtained by just associating  constant ghosts to global symmetries.  This extension, consisting of trivially gauging the global Abelian symmetries, poses no extra obstruction to the solution of the master equation, as it happens in   the case of gauge anomalies.  We illustrate the procedure with the axial model and also calculating the  Adler Bell Jackiw anomaly.    

\vskip 3cm
\noindent PACS: 03.70.+k, 11.10.Ef, 11.15.-q

\vfill\eject

\section{Introduction}
The Batalin Vilkovisky (BV) (or Field-Antifield) method is a Lagrangian  path integral quantization scheme for general gauge theories\cite{BV1,BV2}.
At classical level (zero order in $\hbar$), given a field theory and the associated gauge algebra,  one has a systematic
approach of  building up a gauge fixing structure, even when it involves a chain of ghosts for ghosts, as in the case of reducible theories.
 This procedure generalizes the original idea
of Faddeev and Popov\cite{FP}.  
The condition of BRST\cite{BRST,ZJ}  invariance, at this level , is translated in the so called classical master equation. This equation is mathematically well defined and needs no regularization procedure.
Several important results related to this classical level of the BV  formalism are reviewed in the recent literature\cite{HT,DJ,GPS}.

At the quantum level the situation is different. The  quantum master equation is, in principle, just formal, as it involves the ill-defined $\Delta$ operator associated to the
behavior of the path integral measure. 
The Pauli Villars(PV) regularization procedure was successfully
applied to BV in\cite{TPN}, in such a way that one arrives at a well defined interpretation  to the one loop order equation.
This important step was the starting point for a series of results related to calculating gauge anomalies and Wess Zumino terms at this one loop level, that are reviewed, for example, in \cite{GPS}. 

The question of corrections of loop order higher than one in the BV quantization, where the PV regularization can not be applied,  has been  object of very  
 recent investigations.  
One  proposal for  making sense of higher loop BV is the use of the non local 
regularization, in such a way that the action of the operator $\Delta$ is not singular\cite{Paris}. A different approach  is to translate the master equation in relations that do not involve the operator $\Delta$ and then use  the BPHZ renormalization scheme\cite{DJT}.
Both approaches allow the calculation of gauge anomalies at 
higher loops.

In contrast to this large improvement in people's ability to calculate gauge anomalies using BV quantization, global anomalies are simply ignored if one follows the standard approach.
One normally associates  ghost fields with the parameters 
of  local symmetries. Global symmetries may be described by the introduction of constant ghosts. This kind of approach, with a wide list of important related references can be found, for example, in \cite{PS,DMPW}. In the specific case of the Field Antifield quantization, the introduction of constant ghosts 
in order to derive anomaly free Ward identities for theories with both gauge and global symmetries forming a general algebra was discussed in \cite{BHW1,BHW2}.

If one tries to calculate global anomalies \footnote{We are calling as "global anomalies"  the quantum violation in the conservation of Noether currents, and NOT the breaking of global symmetries, that we assume not to happen} in the field antifield formalism with the aid of constant ghosts one finds
a vanishing result. 
The point is that  anomalies appear in the BV formalism 
multiplied by the corresponding ghosts and integrated over space-time. As global anomalies can be total derivatives \cite{ZJ2},as for example the axial anomaly, proportional to $\,\, \epsilon^{\mu\nu\rho\sigma }\,\,Tr ( F_{\mu\nu}
F_{\rho\sigma})\,\,$ , their contributions to $\Delta S$ could vanish upon space time integrations, or possibly give a constant value, that can be absorbed in the normalization, as in the case of topologically non trivial solutions. Anyway the constant ghosts would not lead to new field dependent terms in $\Delta S$, that could reflect the possible non trivial quantum behavior of the theory regarding the global symmetry.    
Therefore, associating constant ghosts to this symmetries would give no extra contribution to the master equation. As a consequence, the BV generating functional would not  generate the appropriate anomalous Ward identities involving the divergencies of the Noether currents.
Therefore, celebrated results, as the Adler Bell 
Jackiw\cite{ABJ,Ja} anomaly of (non chiral) fermions coupled to gauge fields cannot be calculated in the standard BV framework by just introducing constant ghosts. Although the path integral computation of  global anomalies
was achieved almost two decades ago in refs. \cite{Fuj}, by calculating the regularized Jacobian of the associated local transformations, the incorporation of such a procedure in the BV context  is clearly still lacking. 

Considering some classical action with a global Abelian symmetry  with closed and irreducible algebra, we will enlarge minimally the theory field  space in such a way that a new action will be found, with a larger gauge symmetry.  In this extended action the original global Abelian symmetry is realized locally, but the original theory is recovered at some  gauge. The field-antifield formalism can then be applied in the usual way, but now the one loop order master equation will get a non vanishing  extra contribution associated to the original global anomaly. We will see that 
this contribution will be canceled by an appropriate counterterm, in contrast to what happens with essential gauge  anomalies, where one can not find counterterms that solve the master equation. However,
the counterterms associated with  global transformations will have a non trivial role. 
They will contribute to the Ward identities with insertions of  divergencies of the classically conserved global currents, giving the appropriate one loop anomalous corrections.
    
The article is organized in the following way:
in section (2) we briefly review some results of the BV quantization related to anomalous theories. In section (3) we present our general approach to calculate global Abelian anomalies in BV, illustrating it with the axial model in section (4). The axial anomaly of non chiral fermions coupled to gauge fields, the so called Adler, Bell, Jackiw anomaly, is calculated in section (5). Some concluding remarks are left to section (6).

\section{Anomalies in the standard BV quantization}
 
There are presently in the literature a reasonable amount of reviews about BV quantization and (or) its application to anomalous gauge theories, as \cite{HT,DJ,GPS,H1,TP}.
We will thus present just a brief summary of results to be used in the following sections, stressing some specific points, like gauge fixing.
The quantum action $W[\phi^A , \phi^{\ast}_A ]$ is defined in an enlarged space where the set $\phi^A$ includes the classical fields $\phi^i$  plus all the fields possibly required for gauge fixing: ghosts $\,c^\alpha$ , antighosts, auxiliary fields, ghosts for 
ghosts, ... , and  $\phi^{\ast}_A$ are the corresponding 
antifields, each one with Grassman parity opposite to the corresponding field. 
The generating functional is built up as

\begin{equation}
\label{Vacuum}
Z_\Psi \,[ J ]\, = \int\prod D\phi^A 
exp {i\over\hbar} \left(
W [\phi^A, \phi^{\ast}_A = {\partial\Psi\over \partial \phi^A}] + J_A \,\phi^A \right) 
\end{equation}

\bigskip

\noindent and the expectation value for an operator $X$ is calculated as:

\begin{equation}
\label{EXPEC}
< \,\, X \,\, >_{_{\Psi\,,J\,}} \, = \, \int \prod D\phi^A 
\, X \, exp \left( {i\over\hbar}
W [\phi^A, \phi^\ast_A = {\partial\Psi\over \partial \phi^A} ]+ J_A \, \phi^A \right) 
\end{equation} 

\bigskip 

\noindent The condition that (\ref{Vacuum}) does not depend on the gauge choice (represented by the gauge fixing fermion $\Psi$) when the sources are not present, or when the source term is BRST invariant, is translated in the so called master equation:

\bigskip

\begin{equation}
\label{Master}
<\,{1\over 2}(W,W)\, - \, i\hbar\Delta W\,>_{_{\Psi\,\,,J}}
\,=\,0\,\,,
\end{equation}

\bigskip

\noindent where we are explicitly calling the attention to the fact that this equation comes in as an expectation value.  In Eq. (\ref{Master}) the antibracket is defined as
 $(X,Y) = {\delta_rX\over
\delta\phi^A} {\delta_lY\over\delta\phi^\ast_A}
- {\delta_rX\over \delta\phi^\ast_A}
  {\delta_lY\over \delta\phi^A}$
and the operator Delta as  $\Delta \equiv
{\delta_r\over\delta\phi^A}{\delta_l\over\delta\phi^\ast_A}\;$.
\noindent We are using the de Witt notation of sum  and integration over space-time variables for repeated indices, when pertinent. It is also useful to observe that in fact
Eq. (\ref{Master}) is deduced as an expectation value like (\ref{EXPEC}) and so it is implicitly
assumed that there exist some $\Psi$ that fixes properly
the redundant degrees of freedom associated to gauge invariance. This point will be important in the developments we are going to present in the next sections.

The operator $\Delta$ involves a double functional derivative in the same
space-time point. Therefore, acting on local functionals, it leads to a singular $\delta(0)$. That is why, as we said in the introduction, the master equation at loop order equal or greater than one is just formal and a regularization scheme must be introduced. 
Expanding the quantum action in a power series  in $\,\hbar\,$: 
$\,\,W[\phi^A,\phi^{\ast}_A ] = 
S[\phi^A ,\phi^{\ast}_A ] +
\sum_{p=1}^\infty \hbar^p M_p [\phi^A ,\phi^{\ast}_A ] 
\,\,$
we can write also  the master equation (\ref{Master}) in loop order.
The two first terms are:

\begin{eqnarray}
\label{Master2}(S,S) &=& 0\\
\label{Master3}
(M_1,S) &=& \,i\, \Delta S
\end{eqnarray}

\bigskip

\noindent The zero loop order action $S$ must satisfy the boundary 
condition:
$\,\, S[\phi^A,\phi^{\ast}_ A= 0]= S_0 [\phi^{i} ] \,\,$ where
$\, S_0 (\phi^{i}) \,$ is the original classical action.
Considering an irreducible gauge theory where the original local
symmetries are of the form:

\begin{equation}
\label{LS}
\delta \phi^i = R^i_\alpha\,(\phi )
\theta^\alpha
\end{equation}

\bigskip

\noindent with $\theta^\alpha $  space time dependent parameters,
gauge fixing is obtained by enlarging the field content of the
theory, associating ghosts $\,c^\alpha\,$ to the local parameters,
and requiring that

\begin{equation}
{\delta_r \delta_l \, S \, [ \phi , \phi^\ast ]\over 
\delta c^\alpha \delta \phi^{\ast}_i} 
\vert_{_{\phi^\ast = 0}} =  R^i_\alpha\,(\phi ) \,.
\end{equation}

\bigskip

Eq. (\ref{Master2}) contains all the classical gauge
structure of the original theory. Quantum obstructions of
the gauge invariance may appear at one loop order  and
beyond. As it was already observed, they need to be regularized. The one loop order equation may be regularized using the Pauli
Villars procedure \cite{TPN,TP,DJ,GPS}. When there is no local $M_1$ term
in the original space of fields that solves  eq. (\ref{Master3}), the theory has a gauge (or local)
anomaly. The violation of (\ref{Master3}), for the case of irreducible theories with closed gauge algebra, may be written as

\begin{equation}
\label{Anomaly}
{\cal A }[\,\phi, \phi^\ast \,]\, = \, \Delta S + { i \over  \hbar}
( S , M_1 ) \,=\,a_\alpha\,c^\alpha
\end{equation}

\noindent and translates the possible non trivial behavior of the path
integral measure with respect to some of the local symmetries.
Actually the form of (\ref{Anomaly}) depends both on the
regularization procedure used to calculate $\Delta S$ and on the
counterterms $M_1$ that one is choosing.

\section{Global Abelian Anomalies in the BV Quantization}
\bigskip

Let us consider a classical action  $\,S_0\,[\,\phi^i\,]\,$,  invariant under  local transformations\break 
$ \delta \phi^i = R^i_\alpha [\phi] \theta^\alpha (x)$
and therefore  satisfying the
Noether identities

\begin{equation}
\label{Noether1}
{{\delta S_0}\over{\delta\phi^i}}\,R^i_\alpha\,=\,0\,.
\end{equation}

Besides the local invariances, that we assume to be non anomalous, the action  $S_0$ has possibly a large number of global symmetries also \cite{BHW1,BHW2} . We will not be concerned with the whole set of global transformations
that leave $S_0$ invariant, but just investigate a  particular Abelian subset associated to global anomalies and satisfying  some properties that, as we will  see, will hold for very important cases, like the axial Abelian anomaly.      
Let us split the set of classical fields $\phi^i$ into two subsets $\,\phi^i \,=\,\{A^m\,,\,\psi^r\}$, such that the fields $A^m$ are invariant under the global transformations considered. Thus, writing together the global and local infinitesimal transformations we have

\begin{eqnarray}
\delta A^m&=&R^m_\alpha \,[\,\phi\, ]\,\, \theta^\alpha\, (x)\, , \\
\delta\,\psi^r&=&R^r_\alpha \,[\,\phi \, ]\,\, \theta^\alpha \, (x) 
\,+\, \overline R^{\,r}_a \,[\,\phi\, ]\, \epsilon^a\,\,,
\label{3a}
\end{eqnarray} 

\noindent where $\epsilon^a$ are constant parameters.

We will now assume that the fields $\psi$ transform exponentially. In other words,  the finite version of relation (\ref{3a}) has the specific  form 

\begin{equation}
\label{FT}
\psi^{\prime\,r\,} [\psi,\theta ,\epsilon ] = exp \{ 
i (T^\alpha \theta^\alpha (x) \,+\, { \overline T}^a 
\epsilon^a ) \}^r_s\, \psi^s
\end{equation}

\noindent where it should be noted that we are using the same notation for the parameters $\theta$ and $\epsilon$ as in the infinitesimal case (\ref{3a}) just to simplify the notation.

Transformations like (\ref{FT}) are what occurs, for instance, in QCD if one is concerned only with the phase transformations of the fermionic fields,
and not   with the Poincar\`e transformations  (where global anomalies are absent). In that case the gauge fields
transform only locally while the matter fields transform  locally
and globally under the gauge as well as the global axial symmetries of the action.

We will furthermore  assume that the  global transformations are Abelian.  
That means:

\begin{equation}
\label{disj} 
[ {\overline T}^a \,,\, {\overline T}^b \,] \,=\, 0
\,=\,
[ T^{\alpha}\,,\,{\overline T}^a\,]  \,\,\,\,.
\end{equation}

The infinitesimal expression (\ref{3a}) can, of course, be obtained from (\ref{FT}) by  Taylor expansions in $\theta^\alpha$ and $\epsilon^a$. Thus we identify the generators as

\begin{eqnarray}
\label{3b}
R^r_\alpha&=&{{\partial \psi^{r\prime}}\over{\partial\theta^\alpha}}\vert_{\theta=\epsilon=0}
\,=\,i\,[\,T^\alpha\psi\,]^r \,\,,  \nonumber\\
{\overline R}^r_a&=&{{\partial\psi^{r\prime}}
\over{\partial\epsilon^a}}\vert_{\theta =
\epsilon=0}\,=\,i\, [\,\overline T^a\psi\,]^r \,\,.
\end{eqnarray} 

 At classical level, a gauged version for the global subset of the symmetries appearing in (\ref{3a}) can be written. More precisely, we can redefine the theory in such a way that the set of parameters $\epsilon^a$ becomes  space-time dependent and, at the same time, it is imposed that the old theory is recovered in some gauge. In order to reach this goal we introduce collective fields $\chi^a$ \cite{DNS} and define (see (\ref{FT}))

\begin{equation}
\label{CC}
{\overline \psi}^r \equiv \psi^{\prime\,r} [ \psi , 0,\chi] =
\Big[ exp\{ i {\overline T}^a \chi^a  \}\,\Big]^r_s \psi^s
\end{equation}

By using
(\ref{3a})-(\ref{CC}), we see that $\overline\psi^r$ is invariant under the  transformation associated to the parameter $\epsilon^a(x)$, now made
local, if we impose that the new fields $\chi$ also transform appropriately. The new gauge transformations that leave (\ref{CC}) invariant read:

\begin{eqnarray}
\label{TQ}
\delta\psi^r &=&\,i\, [\overline T^a \epsilon^a (x)\,
\psi ]^r \nonumber\\
\delta\chi^a &=& -\epsilon^a(x)\,\,.
\end{eqnarray}

At this point we extend the  action $S_0[\phi^i]=S_0[A^m,\psi^r]$ to a new action \break $S_1[A^m,\psi^r,\chi^a]$ in such a way that

\begin{equation}
\label{EA}
S_1 [\,A,\psi,\chi] \equiv S_0 [\,A,\overline\psi \, ].
\end{equation}

As $S_1$ depends on $\psi$ and $\chi$ only through
$\overline\psi$, it becomes clear that the action $S_1$ is invariant under
the transformations parameterized by 
$\epsilon^a(x)$. Therefore  

\begin{equation}
\label{delta1}
 \delta \, S_1 [\phi,\chi]_{\vert_{\theta = 0}}\, = 0\,.
\end{equation}

\noindent  We observe that this is true  only because we are assuming that the original global symmetries are Abelian (eqs. (\ref{disj})).

Now, the process of gauging the symmetry associated to the set of parameters $\epsilon^a$ does not modify the form of the generators
$R^m_\alpha$ and $R^r_\alpha$. Then

\begin{equation}
\delta\, S_1 \,[\,\phi ,\chi ]_{\vert_{\epsilon = 0}} =
 {\delta S_0 [\overline\psi , A] \over \delta \overline \psi^r} \,\,
\delta  {\overline \psi}^r\,+{\delta S_0 [\overline\psi , A] \over\delta A^m} \,\,
\delta A^m\,,
\end{equation}

\noindent and  therefore the Noether identities (\ref{Noether1}) of the original local gauge sector continue to be valid.
This means that it is possible to  gauge rigid symmetries of $S_0 [\phi^i\,]$
of the kind expressed by (\ref{FT})  without destroying its original local ones, as long as  (\ref{disj}) holds also.

An important point to be remarked is that  $S_1 [A,\psi,\chi]$ recovers $S_0 [\phi]$ in the gauge $\chi^a = 0$. This assures that the original theory, at least at the classical level, is not changed.  So, we have succeed in
gauging some global symmetries  of a classical action $S_0$, submitted to the quoted restrictions, by enlarging the configuration space in such a way that  the
original theory is recovered when the new
fields are set to zero. Although this is a simple construction, the existence of the new local symmetries of the classical action $S_1$ will enable
a non trivial incorporation of global anomalies in the field antifield
formalism. As we are going to show, this gives us room for introducing  {\it non-constant} ghosts associated to global symmetries. This will be essential in the process of building up anomalous Schwinger-Dyson equations reflecting the non conservation of Noether currents.

As we are assuming that $S_0$ is invariant under the global symmetry  appearing in (\ref{3a}), we have\cite{ZJ2}:

\begin{equation}
\label{QQ}
{\partial {\cal L}_1 \over \partial \chi^a} = 0\,\,,
\end{equation}

\noindent where ${\cal L}_1$ is the Lagrangian density associated to $S_1$. This implies that

\begin{equation}
\label{18}
{\delta S_1 \over \delta \chi^a } \vert_{_{\chi^a = 0}}
= - \partial_\mu J^\mu_a
\end{equation}

\noindent where the current

\begin{equation}
\label{19a}
J^\mu_a = {\partial {\cal L}_1\over \partial (\partial_\mu \chi^a ) } \,\vert_{_{\chi^a = 0 }}
\end{equation}

\noindent is just the on shell classically conserved Noether current. In the derivation of ({\ref{18}) we are assuming
that $S_0$ does not depend on
higher derivatives in $\phi$.
Result (\ref{18})  will  be important in the calculation of the anomalous non conservation of currents (\ref{19a}). 
\bigskip

At this point it is interesting to compare the introduction of the fields $\chi^a$ here with the introduction of fields associated to gauge group elements in order to generate
Wess Zumino(WZ) terms for anomalous gauge theories as it was done first for chiral QCD2 in \cite{BM} and then generalized in \cite{GP}.
In these articles, where true gauge anomalies are present, 
one considers extra WZ fields that are not present at the classical level but will show up just at the one loop order term of the action ($M_1$). It was shown in \cite{JST} that if one  appropriately includes  ghost fields associated to the invariance of the classical action with respect to these extra WZ fields one would get the result that gauge anomalies are never  canceled by the Wess Zumino terms but just shifted to other symmetry.
 
Here we have a completely different situation. Our classical action $S_1$  is NOT independent of the extra fields $\chi^a$. 
Although (\ref{QQ}) may seem to indicate it, we realize by  (\ref{19a}) that $S_1$ depends on $\chi$ through its space time derivatives.
So, we can not transform $\chi$ arbitrarily as it happens with the WZ fields in \cite{BM,GP,JST}. Actually, we can see in (\ref{TQ}) that the new gauge invariance corresponds to a simultaneous change in $\psi$ and $\chi$. We will gauge fix it with extra ghosts, but we do not have a shift symmetry in $\chi$ like those discussed in \cite{JST}.
It is important to note that here, in contrast to\cite{BM,GP,JST},  we are considering  theories with no gauge anomalies. Thus, our procedure for detecting the anomalous violation of Noether currents should not make our theory become gauge anomalous.
We are just building up an enlarged  gauge theory that reproduces the original (non anomalous) one for some partial gauge fixing.
If, as assumed,  the master equation was solvable in the original theory, the same thing is expected to hold in the enlarged theory. The important difference is that, as we will see, we will have to add a counterterm that will generate the Green functions with the insertion of the divergence of the local current considered.

\bigskip

In order to quantize the theory along the field-antifield line, we introduce, besides the usual
ghosts $c^\alpha$ corresponding to the original gauge symmetries of $S_0 [ \phi^i ] $, new 
non constant Abelian ghosts $C^a$  corresponding  to the symmetry (\ref{TQ})
of $S_1 [ \phi^i , \chi^a ]$,
as well as their 
corresponding antifields. The  classical BV action has then the general form:

\begin{eqnarray}
\label{18A}
S [ \varphi^I\,,\,\varphi^{\ast}_I ]  &=&S_1 [ \phi^i , \chi^a ] +
\phi^{\ast}_i R^i_\alpha c^\alpha - c^\ast_\alpha\,
T^\alpha_{\beta\,\gamma}\, c^\gamma\, c^\beta \nonumber\\ &+&
\psi_r^{\ast}{\overline R}^r_a C^a - \chi^\ast_a C^a  + \bar\pi_a\bar
C^{\,\ast\,a} \,\, + ...
\end{eqnarray}
 
\noindent where we have introduced  the notation  $\{\varphi^I\}
\equiv \{ \phi^A,  \chi^a , C^a , \bar \pi^a\,,\, \bar C^a
\}$ for the complete set of fields. In the above equation the $T$'s
are structure constants associated  to the original gauge algebra,
which we are assuming that is closed, irreducible and disjoint.  It
is worth  to mention that as in  eq. (\ref{TQ}) the parameter is
local, the ghosts $C$ (as well as $c$) are also  local, in contrast
to the standard approach of associating constant ghosts to global
symmetries.
\bigskip

Anomalies then formally  come from

\begin{equation}
\label{19}
\bigtriangleup S \, = \, {{\delta R^i_\alpha}\over{\delta\phi^i}}\,\,c^\alpha
\,+\, {{\delta {\overline R}^i_a}\over{\delta\phi^i}}\,\,C^a
\end{equation}

\noindent but this expression is actually ill-defined, as explained in section (2). A  precise  meaning to it can only be given after a regularization procedure is introduced.
We will assume that there is no gauge anomaly in the original theory corresponding to $S_0$ and we will also take as a prescription 
that the first term on the right hand side of (\ref{19}) is to be regularized as in the case without $\chi$. That means that it will contribute at maximum to a trivial (BRST exact) $\,\,\chi\,\,$ independent term that could be absorbed by adding an appropriate counterterm to $S_0$. 
Therefore only the second term on the right hand side of (\ref{19}) will represent relevant contributions, and consequently we will consider the regularized version of (\ref{19}) to be just proportional to the new ghosts $C^a$:

\begin{equation}
\label{RegJac}
(\,\bigtriangleup\,S )_{Reg.}\,  \,=  \, i\,(\,J_a\,)_{Reg.}\,C^a\,.
\end{equation}

The important question at this point is: does (\ref{RegJac})
represent a true gauge anomaly?  In the field antifield quantization
we say that a theory has a gauge  anomaly when, after calculating a
regularized $\Delta S$, one verifies that  the one loop order master equation
(\ref{Master3}) does not admit a local solution in the original space
of fields and antifields. 
In other words, when it is not possible to find a counterterm $M_1$
whose BRST variation is proportional to $(\,\bigtriangleup\,S )_{Reg.}$. From the
cohomological point of view\cite{HT}, an anomaly corresponds to 
$(\,\bigtriangleup\,S )_{Reg.}$ been BRST closed but not BRST exact.
However, from the form of the BRST transformations of $\chi^a$ and
$C^a$ 

\begin{eqnarray}
\label{doublets}
\delta_{_{BRST}} \chi^a &=& - C^a\nonumber\\
\delta_{_{BRST}} C^a &=& 0
\end{eqnarray}

\noindent one realizes that these two fields are absent from the cohomology
\cite{HT} ( they constitute what is called a BRST doublet \cite{PS}).
That means, eq. (\ref{RegJac}) represents just a BRST exact term for
which one can always find a local $M_1$ term that solves the master
equation (\ref{Master3}).

The explicit form of such a counterterm will depend on (\ref{RegJac}) but
it can be put in the form:

\begin{equation}
\label{23}
\hbar \, M_1 \,=\,+ \hbar \chi^a\, (\,J_a\,)_{Reg.}\,\vert_{_{_{\chi\,=\,0}}}
+ O [\, \chi^2\,\,] .
\end{equation}

\noindent where $ O [ \,\chi^2\,\,]$ means terms of order two or more in the $\,\chi^ a\,\,$ fields. 

\noindent Defining the generating functional as:

\begin{equation}
\label{GenGl}
Z_{\,\overline \Psi}\,[ J_A ]\,=\, \int [d\varphi^I]
exp {i\over \hbar} \Big( S_{_\Sigma}\,\,+\,
\hbar M_1 [\phi^A, \chi^a ]\,\,+\,\, J_A\, \phi^A \Big)
\end{equation}

\noindent
were $ S_{_\Sigma} \,=\, S[\varphi^I , \varphi^{\ast}_I = {\partial \overline \Psi \over \partial \varphi^I} ]\,\,$ 
and with gauge fixing fermions constrained to the form:

\begin{equation}
\label{27}
\overline \Psi [\varphi^I ] =\,\bar C_a\,\chi^a\,+\, \Psi [ \phi^A ]
\end{equation}

\noindent the symmetry (\ref{TQ}) is always fixed in the trivial gauge  
$\chi^a\,=\,0$ and the original theory is recovered.

This answers our question about the existence of gauge anomalies. The
theory remains (gauge) anomaly free. This is just what could be
expected if $S_0[\phi]$ and $S_1[\phi,\chi]$ represent the same
theory at quantum level.  However, the extra contribution $M_1$ to
the quantum action has a non trivial role.
As the master equation is satisfied, the change in the functional (\ref{GenGl}) under small changes $\delta \overline \Psi$ in the gauge fixing fermion, for the present case where the classical action is linear in the antifields and $M_1$ does not involve antifields, is\cite{TPN}: 

\begin{eqnarray}
 Z_{\,\,{\overline \Psi}^{\,\,\prime}}\,\,[ J_A ] &-& 
 Z_{\,{\overline \Psi}}\,[ J_A ]\,\,\,=
\nonumber\\ &=&
\int [ d\varphi^I]\,exp \lbrace {i\over\hbar} 
\,\Big(  S_{_\Sigma}\,+\,
 \hbar M_1 [\phi^A, \chi^a ]\,+\, J_A\, \phi^A \Big) \rbrace
\,\,{i\over \hbar} J^A\,{\partial_l S \over \partial 
\phi^{\ast\,A}}\,\,\,{i\over\hbar} \,\delta {\overline \Psi}
\nonumber\\
& &
\end{eqnarray}

We will choose the particular variation:

\begin{equation}
\delta {\overline \Psi} \,=\,{\overline \Psi}^{\prime}
\,-\,{\overline \Psi}\,
 = \,{\overline C}^a \,\epsilon^a
\end{equation}

\noindent where $\epsilon^a\,(x)\,$ are small arbitrary quantities. Furthermore, we will assume that the gauge fixing fermion  does not depend on the fields $\psi^r\,\,$ (this is what  happens, for example, in ref\cite{Fuj}, where the gauge fixing part of the action is  implicitly assumed not to depend on the fermionic matter fields):

\begin{equation}
\label{GF3}
\overline \Psi [\varphi^I ] =\,\bar C_a\,\chi^a\,+\, 
\Psi [ A^m\,, c_\alpha , ... ]
\end{equation}

\noindent where the dots refer  to possible trivial pairs but not to $\psi^r$.

From this condition and (\ref{18A}) we find

\begin{equation}
{i\over \hbar} \overline C^a exp {i\over \hbar} \Big(
S_{_\Sigma} + \hbar M_1
\,+\, J^A \,\phi_A \Big)\,\,=\,\,
{\partial \over \partial C^a }\,  exp {i\over \hbar} \Big(
S_{_\Sigma} + \hbar M_1
\,+\, J^A \,\phi_A \Big)
\end{equation}

\noindent Partially integrating in $C^a$ and using again  
(\ref{18A}) we get

\begin{equation}
\label{V1}
Z_{\,\,{\overline \Psi}^{\,\,\prime}}\,\,[ J_A ]\,-\, 
\, Z_{\,{\overline \Psi}}\,[ J_A ]\,\,\,=
{i\over \hbar}  \int [ d\varphi^I ]\,\,\epsilon^a
J_r\, {\overline R}^r_a \,\, exp {i\over \hbar} \Big(
S_{_\Sigma} + \hbar M_1
\,+\, J^A \,\phi_A \Big)
\end{equation}

On the other hand, we can explicitly calculate
$Z_{\,{\overline \Psi}^{\,\prime}}\,$ and then,
changing variables to $\chi^{\prime\,a}\,=\, \chi^a \,+
\, \epsilon^a\,$ (which has Jacobian one) and Taylor expanding in this variable get:

\begin{equation}
\label{V2}
Z_{\,\,{\overline \Psi}^{\,\,\prime}}\,\,[ J_A ]\,-\, 
 Z_{\,{\overline \Psi}}\,[ J_A ]\,=\,
-{i\over\hbar} \int [d\varphi^I]
\epsilon^a\,{\partial\over \partial \chi^a} \Big(
 S_1 \,+\,\hbar M_1 \Big) 
 exp {i\over \hbar} \Big(
S_{_\Sigma} + M_1
\,+\, J^A \,\phi_A \Big)
\end{equation}
 
\noindent Note that, as in (\ref{GenGl}), integration over the $\overline \pi^a$ fields will give  delta functionals on $\chi^a$ that will remove any possible new interactions involving these extra fields.

From (\ref{V1}) and (\ref{V2}) we get:

\begin{equation}
\label{Expect3}
\int [d\varphi^I]
\,\epsilon^a \, \lbrace\,
\,{\partial\over \partial \chi^a} \Big(
 S_1 \,+\,\hbar M_1 \Big)
\,+\,J^r {\overline R}^r_a \,\rbrace\,
exp {i\over \hbar} \Big(
S_{_\Sigma} + \hbar M_1 
\,+\, J^A \,\phi_A \Big)\,=\,0
\end{equation}

Now using  eqs. (\ref{18}) and (\ref{23}) and considering that (\ref{Expect3}) is valid for arbitrary $\epsilon^a (x)$ we find, after integrating over ${\overline \pi}^a , {\overline C}^a ,C^a , $ and $\chi^a$:

\begin{equation}
\label{Expect2}
\int [d\phi^A]  \Big( \partial_\mu J^\mu_a (x) - 
\hbar (\,J_a\,)_{Reg.}\vert_{_{\chi\,=\,0}}\,(x)
\,-\,J^r {\overline R}^r_a\,(x)\,\, \Big)
exp {i\over \hbar} \Big( S_{BV}  
+\,\, J^A\, \phi_A \Big)\,\,\,=\,\,0
\end{equation}

\noindent where 

\begin{equation}
S_{BV} \,=\,
\,S_{BV}\,[\,\phi^A , \phi^{\ast}_A =
 {\partial \Psi \over \delta \phi^A}\,]\,=\,
S_0 [ \phi^i ] \,+
\,{\partial \Psi \over \partial A^m} R^m_\alpha c^\alpha - {\partial \Psi \over \partial c^\alpha} 
T^\alpha_{\beta\,\gamma}\, c^\gamma\, c^\beta \,
\end{equation}

 \noindent is just the BV action for the original theory, assuming condition (\ref{GF3}) to hold.
So, we obtain the expectation value (in the original space of fields, with no more $\chi$ fields) of the divergence of the Noether's
current.  Expanding (\ref{Expect2})  in the sources $J_A$ we get the whole set of Greens functions involving the insertion of the  operator $ \partial_\mu J^\mu_a $ \cite{PR}.  This corresponds to the results
derived in \cite{Fuj}, of course outside of the BV framework. Therefore, the generating functional (\ref{GenGl}) contains more information than the original BV generating  functional of eq. (\ref{Vacuum}),
that only contains information about Greens functions involving the original vertices of the theory.

\section{Axial Model}
Let us begin by considering the model described by the classical action\cite{RS}:

\begin{equation}
\label{Axial}
S_0 = \int d^2 x \,\Big( i\overline\psi \gamma^\mu \partial_\mu \psi - 
g_0 \overline\psi \gamma^\mu \gamma_5 \psi \partial_\mu \phi 
+ {1\over 2} \partial_\mu \phi 
\partial^\mu \phi - {m^2\over 2} \phi^2 \Big)
\end{equation}

\noindent As action (\ref{Axial}) presents no  gauge invariance, following
standard BV quantization we do not need ghosts at all. The quantum
action, to be used in the generating functional would therefore be
just (\ref{Axial}). This way we would get no information about
global anomalies in the model.  Let us, however, investigate 
the two sets of internal rigid symmetries of (\ref{Axial})

\begin{eqnarray}
\label{l1}
\psi &\rightarrow&  \psi^{\prime} = exp (ie \epsilon^1 ) \,\psi
\nonumber\\
\overline\psi &\rightarrow& \overline\psi^{\prime} =
\overline  \psi\, exp ( -ie\epsilon^1 )
\end{eqnarray}

\noindent and

\begin{eqnarray}
\label{l2}
\psi &\rightarrow&  \psi^{\prime} = exp (ie\gamma_5 \epsilon^2 \,)\, \psi
\nonumber\\
\overline\psi &\rightarrow& \overline\psi^{\prime} =
\overline  \psi \, exp ( ie\gamma_5 \epsilon^2 \,)
\end{eqnarray}

\noindent with $\epsilon^1$ and $\epsilon^2$ constants. Of course, we could consider the whole set of global symmetries of (\ref{Axial}), but
as we are interested in investigating  possible anomalies  associated to transformations  (\ref{l1}) and (\ref{l2}), let us consider only this set. Let us follow the proposal of section (3) and, in order to investigate the quantum behavior of the Noether currents associated to these global transformations, introduce new fields $\chi^1$ and $\chi^2$ in order to gauge  (\ref{l1}) and (\ref{l2}) respectively along the lines described in the previous section. This results in adding to the action  (\ref{Axial}) the term:

\begin{equation}
- e \int d^2x \Big(  \overline \psi \gamma^\mu \partial_\mu\chi^1 \, \psi  +  \overline \psi \gamma^\mu \gamma_5 \partial_\mu\chi^2 \, \psi \Big)\,.
\end{equation}

\noindent As the sum of eq. (\ref{Axial}) and the term above is invariant now 
under (\ref{l1},\ref{l2}) for local $\epsilon^1$ and $\epsilon^2$ once 
$\chi^1$ and $\chi^2$ transform as
$\delta \chi^1 \,=-\epsilon^1$
and$\,\,\delta\chi^2=-\epsilon^2$, we  consider these symmetries
in the usual way in the field-antifield formalism. So we
include the ghosts $C_1$ and $C_2$ corresponding to $\epsilon^1$ and $\epsilon^2$, and to the previous extended action we add the gauge fixing action 

\begin{eqnarray}
S_{gf} &=& \int d^2 x \Big[ ie \Big ( \psi^\ast \psi C_1 
- \overline\psi \overline\psi^\ast C_1 +
 \psi^\ast \gamma_5 \psi C_2 + \overline \psi \gamma_5 \overline\psi^\ast C_2 \big) \nonumber\\
&-& \chi^{  2\,\ast} C_2 - \chi^{1\,\ast} C_1
+ \pi^2 \overline C_2^\ast  + \pi^1 \overline C_1^\ast \Big] \,.
\end{eqnarray}

Following the ideas introduced in the previous section, we can 
investigate the anomalies associated to the gauge symmetries so
introduced. As the quantum master equation is not well defined,
we need to adopt some regularization procedure.
A rich regularization  scheme (at first order in $\hbar\,$) can be given by the Pauli Villars (PV) procedure \cite{DJ,TPN,DTNP}. 
In order to properly implement this kind of regularization,
we need to introduce PV fields with convenient definitions for the path integral in such a way that the whole measure for the PV fields and the the original ones is BRST invariant. The PV action is also constructed in such a way that the only source of BRST non invariance comes from the PV mass term.  If we choose a mass term for the fermionic PV fields that has the usual form as the mass term for Dirac fermions, we can show that after integrating out the PV fields and taking the infinite limit of the regulating mass, we get,

\begin{equation}
\label{DAX} 
(\Delta S)_{Reg.} = - {e\over \pi} \int d^2 x \, C_2\, 
\Big( g_0 \Box \phi + e \Box \chi^2 \Big)\,.
\end{equation}

\noindent This result  corresponds to the usual Fujikawa regularization where the vectorial transformation is considered as a prefered symmetry.
\noindent With this in consideration, we see that the master equation at one loop level (\ref{Master3}) can be satisfied for this $(\,\Delta S)_{Reg.\,\,}$ choosing the local counterterm

\begin{equation}
M_1 = { i e \over\pi} \int d^2 x \Big( 
 g_0 \chi^2 \Box \phi + {e\over 2} \chi^2 \Box \chi^2 \Big)\,.
\end{equation}

This shows that the theory has no gauge anomalies, what should be
expected from the  original theory
described by (\ref{Axial}). The presence of the counterterm $\hbar M_1$ in the quantum action will however enlarge the content of the generating functional, in the sense that it will also allow the calculation of the quantum expectation values of divergencies of the Noether currents. Writing out
equation (\ref{Expect2}) for $\chi = \chi^2$ and then for $\chi = \chi^1$ and taking the zero order term in the sources, we find respectively:

\begin{eqnarray}
\label{Ward}
<\partial_\mu ( \overline \psi \gamma^\mu \gamma_5 \psi )
>_{_{ \Psi\,}} &=& - i {g_0 \over \pi} < \Box \phi >_{_{ \Psi\,}}\nonumber\\
<\partial_\mu ( \overline \psi \gamma^\mu \psi )
>_{_{\Psi\,}} &=& 0
\end{eqnarray}

So the results of ref. \cite{RS}  are reproduced in the field-antifield formalism, once we gauge the axial symmetry in a minimal way.

As a final comment, we observe that we could have chosen other mass terms. For instance, we could have introduced
non transforming PV fields, besides the usual PV fields\cite{TPN} and easily construct a PV mass term 
that would be non invariant under the chiral and the vector symmetries.  Under these scheme, both symmetries would be non preferential and, as a result, the anomaly would appear
along $C_2$ as well as along $C_1$. Because of the cohomological triviality, the master equation would again be satisfied for some $M_1^\prime$ and following the same procedure used in this section, we would generate an  anomalous  divergence for the vectorial current as well.
These results can also appear  in the calculation of anomalous divergencies by following the procedure of Fujikawa.

\section{Adler Bell Jackiw Anomaly }
Let us consider fermions coupled to non-Abelian gauge fields, described by the action:

\begin{equation}
\label{LSY}
S_0 = \int d^4 x \Big[ - {1\over 4} Tr ( F^{\mu\nu}\,
F_{\mu\nu} ) + i\overline \psi \gamma^\mu \big(
\partial_\mu -ig A_\mu \big) \psi \Big]
\end{equation}

\noindent with $A_\mu = A_\mu^\alpha T^\alpha$ and $[ T^\alpha , T^\beta \,]\, = i f_{\alpha\beta\gamma} T^\gamma $. This action is invariant under the local infinitesimal transformations:

\begin{eqnarray} 
\label{GSY}
\delta \psi &=&  ig\omega^\alpha (x)\, T^\alpha \psi 
\nonumber\\
\delta \overline \psi &=& - ig \overline \psi \omega^\alpha (x)\,T^a\, 
\nonumber\\
\delta A_\mu &=& D_\mu \omega
\end{eqnarray}

\noindent If we were to follow the steps of the standard BV quantization, we should include just one (non Abelian) ghost, say $c\,=\, c^\alpha\,T^\alpha$, associated to (\ref{GSY}). However, we want to investigate the behavior at the quantum level of the  global Abelian axial symmetry of (\ref{LSY}):

\begin{eqnarray} 
\psi^{\prime} &=& \psi_\beta = e^{ig\beta} \psi 
\nonumber\\
\overline \psi^{\prime} &=&\overline \psi_\beta =  \overline \psi e^{- ig\beta} 
\nonumber\\
A_\mu^{\prime} &=& A_{\mu} 
\end{eqnarray}

Thus, following the steps of the previous section, we introduce a bosonic Abelian field $\chi$ and an Abelian ghost $C$ associated to  the global symmetries 
(\ref{GSY}), writing the total action:

\begin{eqnarray}
\label{QCDA}
S &=&  \int d^4 x \Big[ - {1\over 4} Tr ( F^{\mu\nu}\,
F_{\mu\nu} ) + i\overline \psi \gamma^\mu \big(
\partial_\mu -ig A_\mu  + ig \gamma^5 \partial_\mu \chi\, \big)\, 
\psi \nonumber\\ &+& A^{\ast\,\alpha}_\mu \,(D^\mu c)^\alpha 
- ig \psi^\ast c \psi  + ig\overline\psi \,c \overline\psi^\ast  + 
ig \psi^\ast \gamma^5 \psi C + ig \overline \psi \gamma^5
\overline \psi^\ast  C \nonumber\\
 &+& {1\over 2}  c^{\ast\,\alpha}\,c^\beta\,c^\gamma f_{\alpha\beta\gamma} 
- \chi^\ast C + \overline \pi \,\overline C^\ast \Big]
\end{eqnarray}

The next step is to calculate $\Delta S$ for this enlarged action that includes, besides the original gauge
coupling, an additional axial  one. This situation, in the field antifield context, was considered in \cite{TPN}, 
where the results were shown to be equivalent to previous calculations presented in \cite{HYM} for the fermionic Jacobian, in this case of mixed coupling, using Fujikawa's regularization.
Considering, as is the case here, that the axial field is Abelian, their result simplifies to:

\begin{eqnarray}
\Delta S_{Reg.} &=&  - {g^3 \over 16\pi^2} \int d^4x C \epsilon^{\mu\nu\rho\sigma} 
\,Tr\, F_{\mu\nu}\,F_{\rho\sigma}
 \,-\, \Lambda^2\, {g^2 \over 2 \pi^2 } \int d^4x\,C\, \Box \chi  \nonumber\\
&+& {g^2\over 12\pi^2}\int d^4x\,\,C\,\, \Big( { 1\over 6}
\Big( \Box^{\,2} \chi  + 4 g^2 \,(\,\partial_\mu \chi \,\Box
\chi \,\partial ^\mu \chi \,+ 
2 \partial_\mu\partial_\nu \chi \,\partial^\mu \chi \,\partial^\nu \chi \,\,)\,\Big)\nonumber\\
& & 
\end{eqnarray}

\noindent where $\Lambda^2$ is a regulating parameter.
The $ M_1$ term that solves the master equation in this case is: 

\begin{eqnarray}
M_1 \,&=&\,+ { i g^3 \over 16\pi^2 } \int d^4x \,\chi \epsilon^{\mu\nu\rho\sigma} \, F_{\mu\nu}\,F_{\rho\sigma}
\,+\, {ig^2 \over 4\pi^2}  
\int d^4x \Big( \Lambda^2 \chi \Box \chi - {1\over 6} \chi \Box^{\,2} \chi\Big) \nonumber\\
&-& i {g^4 \over 12\pi^2} \int d^4x \Big( \chi \partial_\mu \chi \Box\chi \partial^\mu \chi + 2 \chi\, \partial_\mu \partial_\nu \chi \,\partial^\mu \chi\, \partial^\nu \chi\, \Big)
 \end{eqnarray}

\bigskip 

The Greens functions with the insertion of $\partial^\mu J_{\mu\,5}\,$ can then be calculated from eq. (\ref{Expect2}), considering a source term like :

\begin{equation}
J^A \phi_A \,=\, \overline\psi \eta \,+ \,\overline \eta \psi
\,+\, J^\mu A_\mu
\end{equation}

\noindent and then expanding in the sources.

One example is:

\begin{eqnarray}
< \partial^\mu J_{\mu\,5} (x) \psi (y) {\overline \psi} (z) \,>_{_{ \Psi\,}} &=&
<  \hbar \,{ig^3 \over 16 \pi^2}\,\epsilon^{\mu\nu\rho\sigma} \, F_{\mu\nu}\,F_{\rho\sigma}(x) \psi (y) {\overline \psi} (z) >_{_{ \Psi\,}} \nonumber\\ &+& i g \delta (x-y) < \gamma_5\psi(y)
{\overline \psi} (z) >_{_{ \Psi\,}}
+ i g \delta (x-z) < \psi(y)
{\overline \psi} (z)\gamma_5 >_{_{ \Psi}} \nonumber\\
\end{eqnarray}

\noindent as in  references \cite{ABJ,Fuj}.

\section{Conclusions}
Global Anomalies play a very important role in the description of processes like the $\,\pi_o$ decay\cite{Ja}.
In this way it is interesting to build  up a generating functional that describes this kind of behavior.
In the standard formulation, the Noether currents associated to global transformations are in general not present in the Lagrangian, in contrast to the local currents, which are  coupled to the gauge fields. That is why the standard BV, in general, does not allow computations of Greens functions involving global currents. 
We have shown in this article that for global Abelian
symmetries, the generating functional in the BV quantization procedure can be built up with extra fields, together with  associated extra gauge degrees of freedom, in such a way that
global  anomalies naturally arise from the generating functional.
It is important  to mention that our approach has some similarities with the field space enlargement used in ref. \cite{AD}. There, the gauge symmetry group of some theory is trivially extended and then an appropriate gauge fixing of the extra symmetries leads to the BV action with an interesting  interpretation for the antifields. 

Our  results were obtained for the case of global Abelian symmetries. Also only irreducible theories with closed algebra
were considered.
More general situations are under study and the results will be reported elsewhere.

\vskip 1cm
\noindent {\bf Acknowledgment:}We are very grateful to M. Henneaux for showing us that the new fields are absent from the
cohomology and also for other very useful discussions. We   
would like to thank also J. Barcelos-Neto, C. Farina de Souza and E. C. Marino
for important discussions. This work is supported in part by
 CNPq, FINEP and FUJB (Brazilian Research Agencies).  

\end{document}